# PANORAMIC SPECTROSCOPY OF GALAXIES WITH STAR-FORMATION REGIONS. A STUDY OF SBS 1202+583

S. A. Hakopian,[1] S. K. Balayan,[1] S. N. Dodonov,[2] A. V. Moiseev,[2] and A. A. Smirnova[2]

**Abstract.**
*The methods of panoramic (3D) spectroscopy are used by us in a detailed study of galaxies with ongoing star formation chosen from among objects in seven selected fields of the Second Byurakan Survey (SBS). This article deals with the irregular galaxy SBS 1202+583, which our classification scheme identifies as being in a continuous phase of star formation. Observations were made with the panoramic spectrographs MPFS at the 6-m telescope of the Special Astrophysical Observatory (SAO) of the Russian Academy of Sciences and VAGR at the 2.6-m telescope of the Byurakan Astrophysical Observatory (BAO) in Armenia. The data are used to construct maps of the radiative fluxes in the continuum and various emission lines. Special attention is devoted to analyzing the emission in the H$\alpha$ hydrogen recombination line and in the forbidden low-ionization doublets of nitrogen [NII] $\lambda\lambda$6548, 6583 and sulfur [SII] $\lambda\lambda$6716, 6731, and the ratios of the intensities of the forbidden lines to H$\alpha$. The observable characteristics (size, H$\alpha$ fluxes, etc.) of nine HII regions are studied. The estimated current rates of star formation in the individual HII regions based on the H$\alpha$ fluxes lie within the range of 0.3-1.2 $M_\odot$/year. The dependence of the ratio of the intensities of the emission in these above mentioned forbidden doublets on the rate of star formation in the HII regions is found.*

Keywords: *3D spectroscopy: H$\alpha$ emission: HII regions— individual: SBS 1202+583*

## 1. Introduction

The Second Byurakan Survey (SBS) [1], which was carried out using the 1-m Schmidt telescope with an objective prism, has been one of the most productive of the low-dispersion surveys for discovering active galaxies.

---

(1) V. A. Ambartsumyan Byurakan Astrophysical Observatory, Armenia; e-mail: susannahakopian@yahoo.com
(2) Special Astrophysical Observatory, Russia

The SBS has a special role in the detection of galaxies with not only distinct signs of star formation activity, such as blue dwarf galaxies (BCG, or blue compact galaxies), as known, but also for finding galaxies with less distinct signs of star formation [2]. This is conditioned, in particular, by the aim to avoid missing the objects being sought while working with the low-dispersion spectra. Thus, the lists included not only objects with obvious indications of correspondence to the selection criteria, but also those with dubious indications. This applied most often to objects with visible magnitudes close to the sample limit. $18^m.5$-$19^m.5$. So it is possible to analyze or work with samples in individual fields of the SBS only after "cleaning up" the original lists by means of follow-up spectroscopy of higher resolution.

We have made slit spectroscopic observations of ~500 objects from the SBS which form the base sample [3 and references therein]. This included galaxies with extended morphologies that were chosen as active from seven selected fields of the SBS [4]. The objects were classified according to our scheme for spectral classification [5], as adapted to the available data and standard classification criteria. This yielded 345 galaxies (more than 70% of the sample) with confirmed signs of star formation activity. Most of the remainder manifested signs of being active galactic nuclei (AGN).

Galaxies with star formation activity are denoted by SfG (Star forming Galaxies) in our classification scheme [5]. For simplicity, the sample is divided into two subclasses, depending on the equivalent width of the H$\alpha$ emission line (by analogy with Ref. 6), which are SfGcont (cont = continual) for $EW(H\alpha) < 100$ Å, and SfGneb (neb = nebular) for $EW(H\alpha) > 100$ Å, which includes the objects with greater current star formation: blue compact galaxies, HII galaxies, etc. Further development of this scheme is one of our priorities. In fact, as opposed to galaxies with nuclear activity, for which a fairly precise classification scheme exists, various authors use different terminology for galaxies with star-formation activity, which can lead to substantial errors, especially in statistical studies employing published data from different sources.

At this point we are making detailed studies of individual galaxies from the SfG sample beginning with the objects that are the most complicated from the standpoint of morphology. These studies are based on panoramic (3D) spectroscopy with the MPFS spectrograph on the 6-m telescope at the Special Astrophysical Observatory (SAO) of the Russian Academy of Sciences and the VAGR spectrograph on the 2.6-m telescope at the Byurakan Astrophysical Observatory (BAO). These data can be used for simultaneous examination of both the spectral and morphological features of the objects and contain much useful information on spectrophotometric properties of galaxies. Besides obtaining unique material on individual objects, we plan, as data are accumulated, to gain a more profound understanding of the nature of star-formation activity in the overall chain of evolutionary processes taking place in the universe.

This article is the second in a series of planned publications on this topic. The first was devoted to SBS1533+574, a two-component galaxy in the nebular phase of star formation [7]. The galaxy SBS1202+583, which is the subject of this article, is in the continuous phase of star formation and has an irregular structure.

This article consists of seven parts, including the *Introduction* and *Conclusion*. Data from the Sloan Digital Sky Survey (SDSS), release DR7 [8], are presented in part 2 and serve as a kind of morphological guide for the subsequent analysis. The third and fourth parts mostly contain information on the making of the observations. Part 5 contains the results of an analysis of the maps of the H$\alpha$ emission intensity. The existence of nine HII regions is demonstrated and their characteristics are reported. Data on emission in the forbidden low-ionization doublets of nitrogen [NII] $\lambda\lambda$6548, 6583 and sulfur [SII] $\lambda\lambda$6716, 6731 are presented in the next to last part 6, followed by the conclusion.

In a later paper we shall examine the kinematics of the ionized gas in this galaxy.

## 2. Morphological features of SBS 1202+583 according to the SDSS

The galaxy SBS 1202+583 (alternatively, UGC 07070 NED02, VV270ab, etc.) was included in the SBS catalog as a binary object, based on the appearance of only emission lines, i.e., without any signs of an ultraviolet excess. Its major characteristics according to data from NED [9] and LEDA [10], i.e., its diameter, visible and absolute B magnitudes, distance calculated with a Hubble constant $H_0$ = 73 km/s/Mpc, and projected scale, are 1'.1×0'.7, 15$^m$.53, -17$^m$.93, 37.2 Mpc, 180 pc/arcsec, respectively.

In the optical images, this galaxy has an irregular structure consisting of individual condensations, which can be seen, especially, in the SDSS DR7 images (Fig. 1a[1]). This largest of the seven comparatively bright condensations is nominally at the center and is denoted by "C" in the figure. The others are designated by the quadrant relative to C in which they lie. All seven are photometric SDSS objects identified by coordinates which correspond to the centers of the circles indicated in Fig. 1a. In this paper they show up as SDSS components. In the SDSS photometric catalogue, component C corresponds to the identification number 587731891652198444, while the identification numbers of the other condensations vary only in the last two digits, which are indicated in Fig. 1a.

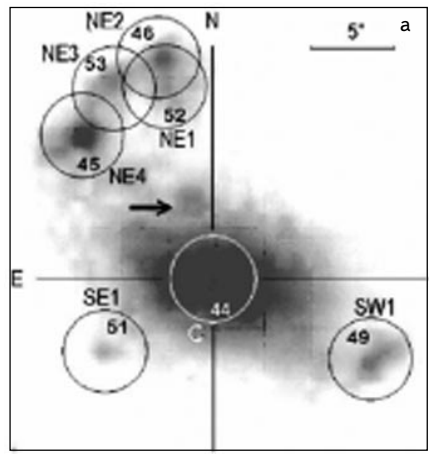

Fig. 1. From the SDSS archive: (a) the sum over five filters of an image of SBS 1202+583 with our notation for the SDSS components and the two last digits which differentiate their identification numbers. The entire number for C is 587731891652198444. (b) The spectrum of the central condensation C and a fragment of an SDSS image with the central component C on a different intensity scale.

---

[1] In the images shown in this article, as in Fig. 1a, north is upward and east is to the left.

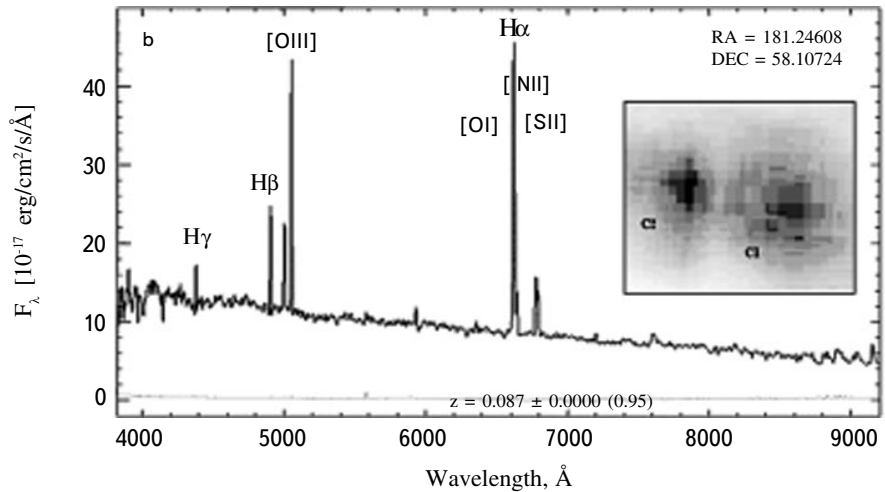

Fig. 1. Continued.

**3. Panoramic spectroscopy: observations and data processing**

SBS 1202+583 was observed using the MPFS (Multipupil Fiber Spectrograph) [11] at the primary focus of the 6-m telescope at the SAO and the VAGR [12] at the primary focus of the 2.6-m telescope at the BAO. The object was observed twice with the MPFS, with different CCD arrays and diffraction gratings, and twice with the VAGR using broad-band interference filters for different spectral ranges. Information on the observations is listed in Table 1. The centers of four fields are shifted relative to one another, but all four contain the SDSS component C. Confirmation of the presence of the SDSS components indicated in Fig. 1a in each of the fields is provided by Fig. 2.

For initial processing of the MPFS observations, involving a standard set of procedures, which include wavelength and flux calibrations, and, partially, for visualization and analysis of the data, there were used programs [e.g. 13] written in Interactive Data Language (IDL). The ADHOCw program package [14] was used for visualizing, analyzing, and illustrating the data obtained with both spectrographs. It also was used for primary processing of the data from the VAGR spectrograph. Here the data were not calibrated in terms of flux because there were no observations of a spectrophotometric standard. The parameters of the lines were determined using a gaussian approximation for the spectral profiles.

TABLE 1. Details on the Observations of SBS 1202+583

| Telescope (observatory) | ZTA 2.6-m (BAO) | | BTA 6-m (SAO) | |
|---|---|---|---|---|
| Spectrograph | VAGR | | MPFS | |
| CCD array | Lick | | Tektronix | EEV 40-42 |
| number of pixels | 2063 × 2058 | | 1024 × 1024 | 2048 × 2048 |
| Field of view (arcsec) | $D \approx 36$ | | 15 × 16 | 16 × 16 |
| Spatial sampling (arcsec) | 0.9 | | 1 | 1 |
| Diff. grating (grooves/mm) | 600 | | 1200 | 1200 |
| Dispersion (Å/pixel) | 2.1 | | 1.35 | 0.76 |
| Observation date | 21 VI 2004 | 15 XII 2005 | 14 V 2002 | 04 XII 2007 |
| Image quality | 2".5 | 3" | 1".7 | 1".9 |
| Spectral range (Å) | 6400-6800 | 4900-5200 | 6000-7300 | 6300-7800 |
| Total exposure (s) | 1800 | 4800 | 1800 | 3600 |
| Spectrophotometric standard | - | - | BD+75d325 | GRW+70d5824 |

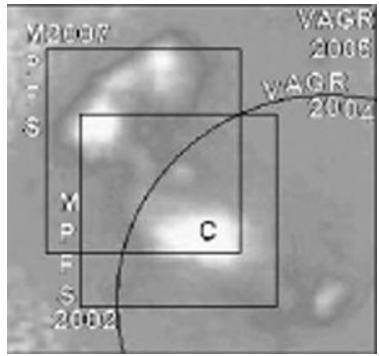

Fig. 2. The sum over five filters of an SDSS image of SBS 1202+583, together with the entire field for our observations with MPFS (indicated by the inner rectangles) and the informative part of the field for the observations with VAGR (outer rectangle and part of a circle).

## 4. Results of observations with the VAGR spectrograph in 2005

All seven of the SDSS components (Fig. 2) show up only the field observed with the VAGR spectrograph in 2005. Only the forbidden oxygen [OIII] λλ5007, 4959 doublet fell within the spectral range of the bright emission lines, i.e., the nebular N1 line and, along the very edge of the range, N2. Figures 3, a, b, and c show maps of the emission derived from the VAGR 2005 data over the entire range of the filter, in a small segment of the continuum, and in the nebular N1 line, respectively. In these figures, the same segment of the sky is isolated as in Fig. 1a, with which Fig. 3a is easily identified. The continuum intensity distribution of Fig. 3b has a single peak.[2] It lies within the confines of the central component and the array element corresponding to it in Figs. 3, b and c, is indicated by the letter "c."

---

[2] Here and in the following, the intensity peak is taken to be the element of the array at which the highest intensity is reached.

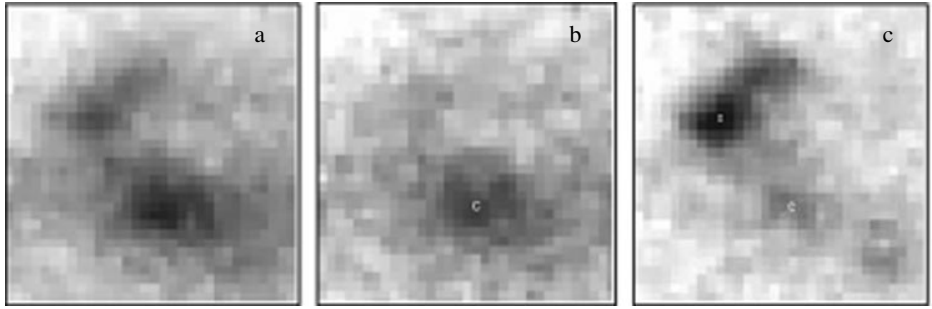

Fig. 3. Distributions of emission intensity obtained with the VAGR spectrograph in 2005: (a) total over the entire spectral range of the filter; (b) total over the continuum segment 5080-5110 Å, for which the peak corresponds to pixel "c" of the array; and, (c) in the [OIII] λ5007 oxygen line. The size of the field is ~30"×30".

The faintest of the SDSS components, SE1 does not stand out in the field. The north-east components NE4 and NE2 of Fig. 3c are the most intense emitters in the nebular doublet lines. Here the highest intensity in the field, which is detected in the peak of component NE4 and is indicated by a cross in Fig. 3c, is almost twice as intense as the emission in the peak from component NE2. In both components the maximum is reached after a smooth rise in intensity from the edge to the center, so that randomness in the position of the peaks can be excluded.

## 5. Recombination line emission. HII regions

The star-formation activity of galaxies originates in their star-formation regions, which are associated primarily with regions of ionized hydrogen, i.e., HII regions. The most distinctive spectral feature of HII regions in the visible range is their emission in the hydrogen Balmer Hα line.

The Hα emission of this galaxy shows up in the three recorded spectral regions for the MPFS 2007, MPFS 2002, and VAGR 2004 observations (Table 1). The two-dimensional distributions of the Hα emission intensity derived from these data are shown as a background in Figs. 4a, 4b, and 5, respectively. With the help of the three partially overlapping fields nine regions of ionized hydrogen were revealed; hence the object under study can be characterized as a complex of HII regions. Six of the nine HII regions are identified with SDSS components.

We use the designations NE5, NE6, and NE7 for the three HII regions that are not associated with SDSS components; they lie to the north east of the central component C. All three are clearly visible in the Hα intensity maps obtained with the MPFS (Figs. 4, a and b). They were also resolved in the VAGR 2004 field, although they lie on the very edge (Fig. 5b). NE7 has the smallest size of all the nine objects. NE5 shows up distinctly in the VAGR 2004 field and in the SDSS images, and is indicated by an arrow in Fig. 1a. It stands out for its small dimensions and circular outlines, and, especially, for the fact that it is observed in all of our fields; this makes it convenient for matching when comparing the data.

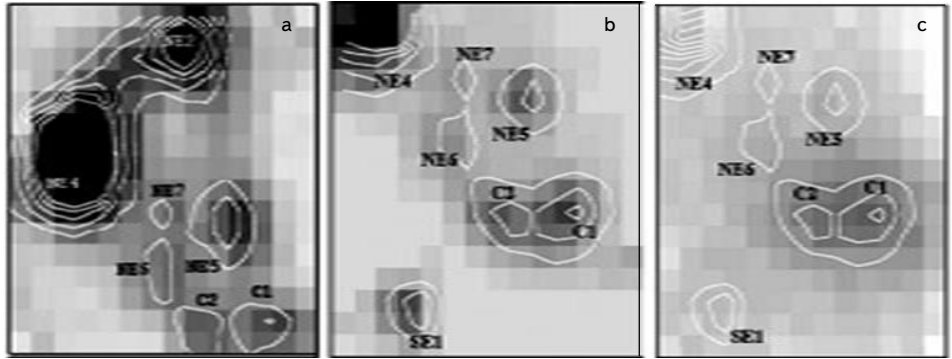

Fig. 4. Flux maps derived from MPFS data: (a) data from 2007; (b, c) data from 2002. The contours[3] in (a), (b), and (c) are drawn in accordance with the Hα emission. The background in (a) and (b) shows the distribution of the radiation in the Hα line, and in (c), the distribution within a continuum segment 6700-6715 Å of the spectrum.

As all three Hα intensity maps show, two HII regions can be observed within the central SDSS component C. The smooth drop in the intensity from the two local peaks observed in C is best illustrated (cf. Fig. 4b) by the white contours in the field of MPFS 2002, where C lies at the center. The binary structure of C also shows up in the SDSS images (see the corresponding fragment in Fig. 1b). We denote the western, larger HII region by C1 and the other, by C2. The continuum emission map derived from the MPFS 2002 data (Fig. 4c) confirms (see part 4, Fig. 3b) the existence of a single peak in this distribution and shows more clearly that it lies in C1 and is coincident with the Hα emission peak of C1.

The HII regions associated with the SDSS components NE4 and NE2 emit more brightly than the others in Hα; these have already recommended themselves as the brightest regions for emission in the nebular N1 line (part 4, Fig. 3c).

Of the recombination lines (besides Hα), only in the spectra of NE4 and NE2 is the singly ionized helium HeII λ6678 line observed. The emission intensity in this line is an order of magnitude smaller than in Hα; at the peak for NE4 it equals $I(\mathrm{He})_{\mathrm{maxNE4}} = 4.34 \times 10^{-17}$ erg/cm$^2$s and it is half this for NE2. The intensity of the Hα emission is given in Table 2.

The MPFS 2007 data, have higher spectral resolution the VAGR data, and in the field of which more HII regions are observed than in the MPFS 2002 and VAGR 2004 data, provide us with homogeneous data on all the seven HII regions mentioned above. These include the largest and brightest of all the HII regions that we have observed. Thus, the range of variation of some of the parameters can be used to characterize the entire set of data as a whole. It should be noted that lower bound estimates are given for regions C1 and C2, since roughly 20-30% of their area is cut off by the field of MPFS 2007, as a comparison with the MPFS 2002 field shows.

---

[3] Here and in the following, if not otherwise specified, the contours reflect an incomplete range of intensities, and the background shading, a complete range; here the darker shades correspond to a higher intensity.

TABLE 2. Comparative and Absolute Characteristics of the HII Regions According to the MPFS 2007 Data

| | NE4 | NE2 | C1 | C2 | NE5 | NE6 | NE7 |
|---|---|---|---|---|---|---|---|
| $I(H\alpha)_{max} 10^{-16}$ (erg/cm²s) | 23.34 | 11.33 | 3.61 | 2.81 | 4.56 | 2.53 | 2.69 |
| $V_r(H\alpha)_{max}$ (km/s) | 2556 | 2554 | 2529 | 2544 | 2548 | 2554 | 2555 |
| $R_{eq}$ (pc) | 460 | 325 | 166 | 145 | 237 | 168 | 84 |
| $F(H\alpha) 10^{40}$ (erg/s) | 14.8 | 9.36 | 5.06 | 4.56 | 5.65 | 4.33 | 4.67 |
| $SFR(H\alpha)$ ($M_\odot$/year) | 1.17 | 0.74 | 0.40 | 0.36 | 0.45 | 0.34 | 0.37 |
| $I([NII]\lambda 6583)_{max} 10^{-16}$ (erg/cm²s) | 0.99 | 0.34 | 0.51 | 0.68 | 0.35 | 0.45 | 0.26 |
| $I([SII]\lambda 6716)_{max} 10^{-16}$ (erg/cm²s) | 1.45 | 0.64 | 0.62 | 0.53 | 0.34 | 0.40 | 0.18 |
| $I([ArIII]\lambda 7136)_{max} 10^{-16}$ (erg/cm²s) | 0.75 | 0.46 | - | - | - | - | - |

The parameter values in the first five rows of Table 2 obtained from data from MPFS 2007 for the seven HII regions are related to the H$\alpha$ emission. In order, these are the peak H$\alpha$ emission $I(H\alpha)_{max}$, the heliocentric radial velocity $V_r(H\alpha)_{max}$ determined from the H$\alpha$ line in the same array element, the equivalent radius of the HII region estimated using the formula $R_{eq} = (S/\pi)^{0.5}$, the integrated radiation flux $F(H\alpha)$, and the star formation rate $SFR(H\alpha)$ calculated from the H$\alpha$ luminosity using the formula $SFR(H\alpha)$ second term in second parentheses:

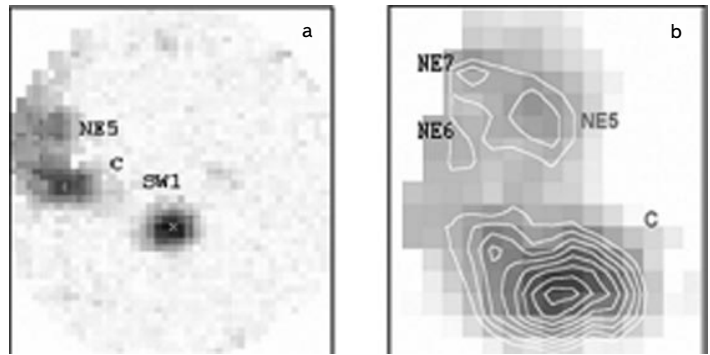

Fig. 5. Distribution of H$\alpha$ radiation, background, and contours derived from the VAGR 2004 data covering the entire field in (a), and the north eastern part of the field (magnified) in (b).

$(M_\odot/\text{year}) = 7.9 \times 10^{-42} L(\text{H}\alpha)$ [15]. The area $S$ of the radiating surface was determined from the number of array elements in which the intensity exceeds a threshold value $I(\text{H}\alpha)_{thresh}$, which, in turn, was determined separately for each HII region, generally using the formula $I(\text{H}\alpha)_{thresh} = 0.1 \times I(\text{H}\alpha)_{max}$.

The table shows, in particular, that the star formation rate of the individual HII regions ranges from a minimum of 0.34 $M_\odot$/year for NE6 to a maximum of 1.2 $M_\odot$/year for NE4. The ranges of the other parameters, except $V_r$, are given by the values for the HII regions NE7 and NE4.

The two HII regions that were not in the field of MPFS 2007 are characterized by intermediate values of the parameters. One of them is associated with the SDSS component SE1 (Fig. 4b) and the other, with SW1 (Fig. 5). Both lie to the south of the central component and, while in projection against the celestial sphere, they are its nearest neighbors, the two appear to be more isolated compared to the other HII regions.

In terms of its parameters SE1 is comparable to NE5. This applies to its shape as well as to the peak intensity and the equivalent radius, which are roughly 1.2 times the values for NE5 according to the data from MPFS 2002.

The HII region SW1 appears in the central part of the VAGR 2004 field (Fig. 5a). The peak of the continuum emission coincides with the local peak of the H$\alpha$ emission for C1; it is indicated by "c." The H$\alpha$ intensity in the peak of SW1 is roughly the same as in the peak of C1, greater by a factor of 1.2, and is greater by a factor of 1.7 than in the peak of NE5. It should be noted that here, closer to the edges of the field where C and NE5 lie, the errors are greater than in the center of the field (where SW1 lies).

## 6. Forbidden line emission of the HII regions according to the MPFS 2007 data

Within the spectral range observed by MPFS 2007 (Table 2), the doublet lines of singly ionized nitrogen [NII] $\lambda\lambda$6548, 6583 and sulfur [SII] $\lambda\lambda$6716, 6731 and the doubly ionized argon [ArIII] $\lambda$7136 line can be seen. The [NII]

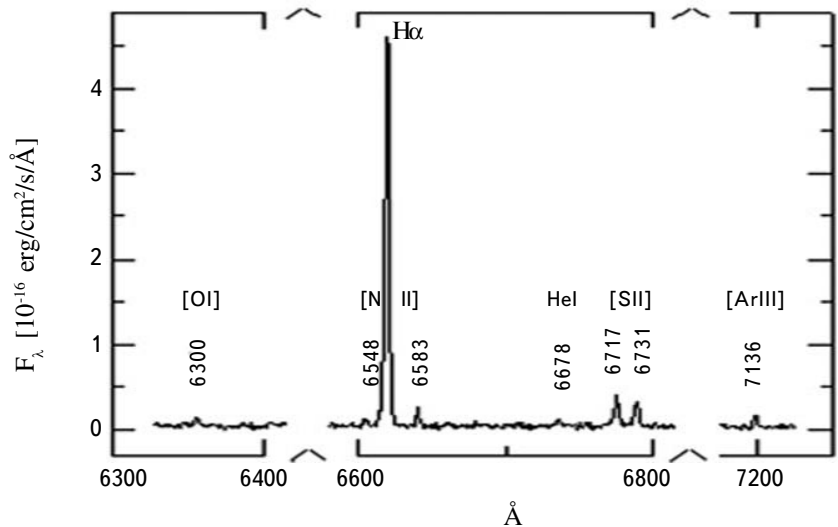

Fig. 6. The three most informative parts of the spectrum, integrated over the area of NE4.

λ6548 line is faint and in most of the spectra its intensity does not exceed the 3σ noise level of the continuum. The picture is roughly the same for the neutral oxygen [OI] λ6300 line, although there are some problems with correct background subtraction for it, since, because of the redshift, $z \approx 0.0087$, it partially overlaps the [OI] λ6363 airglow line. Figure 6 shows the most informative parts of the spectrum obtained by integrating the data over the pixels covering the surface of NE4. All of the emission lines are seen, including the two faint forbidden lines and the HeI λ6678 recombination line.

The distributions of the intensities in the nitrogen [NII] λ6583 line and in the sum of the sulfur [SII] λλ6716, 6731 doublet lines are shown, respectively, in Figs. 7a and 7b by white contours. A comparison of these distributions with the distribution of Hα, which is shown in both figures by dark contours, reveals some differences that are most obvious within NE4. The intensities of the emission at the peaks of the seven HII regions in the three forbidden lines $I([NII]\lambda 6583)_{max}$, $I([SII]\lambda 6716)_{max}$, and $I([ArIII]\lambda 7136)_{max}$ are listed in the bottom rows of Table 2.

The ratios of the intensities of the forbidden lines to the permitted lines (denoted by $R_p^f$ below) are customarily used as a criterion for determining the type of activity in galaxies [e.g. 16 and the references therein]. The ratios $R_p^f$ play a basic role in our classification scheme [5], which was developed for uniform processing of spectral data, especially for the purpose of distinguishing galaxies with nuclear (AGN) or star formation (SfG) activity, even when these indicators are faint. In studying SfG galaxies we are, in particular, interested in a range of values of $R_p^f$ that is determined primarily by the chemical composition of individual HII regions.

The galaxy SBS λ1202+583 being studied here, with its clearly distinct HII regions, is most suitable for this purpose. The uniformity of the spectra data on the seven HII regions observed in the MPFS 2007 field makes it possible to analyze the distribution of the ratios of the intensity of the [NII] λ6583 line and of the sum of the [SII] λλ6716, 6731 sulfur doublet lines to the intensity of the Hα line, which are denoted by $R_I^1 = [\text{NII}]\lambda 6583/\text{H}\alpha$ and $R_I^2 = [\text{SII}](\lambda 6716+\lambda 6731)/\text{H}\alpha$. The distributions of $R_I^1$ and $R_I^2$ over the field of MPFS 2007 are shown in gray-scale in Figs. 7a and 7b; there the scale indicates their range of variation and the numbers given on the figures were obtained by averaging over each of the HII regions. These same values are listed in Table 3, right after the similarly averaged values of the Hα emission intensity.

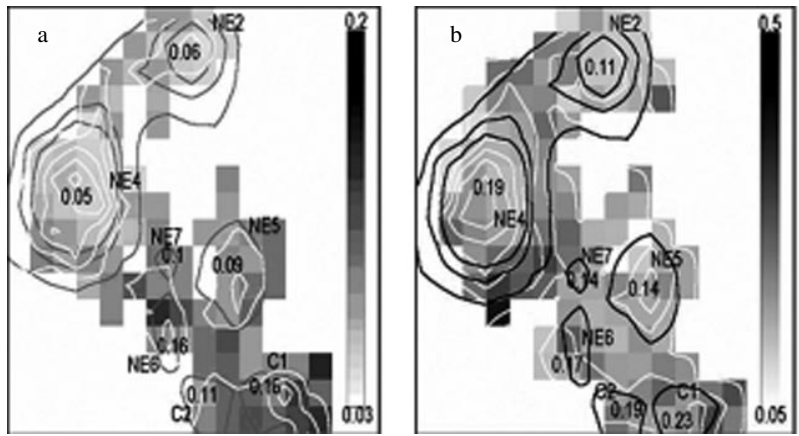

Fig. 7. Distributions derived from the MPFS 2007 observations. The dark contours correspond to Hα emission intensities and the light contours, to the emission intensities in (a) the [NII] λ6583 line and (b) the sum of the [SII] (λ6717+λ6731) doublet lines. The gradations in the gray and the numbers correspond to the line ratios [NII] λ6583/Hα in (a) and [SII] (λ6717+λ6731)/Hα in (b).

The average values of $R_I^1$ for the HII regions lie nonuniformly between 0.05 and 0.16 and are grouped by two each at the beginning and end of this interval, and by three near 0.1, which can be regarded as the maximum of the distribution of $R_I^1$. The maximum of the distribution of the averages of $R_I^2$ for the HII regions lies at 0.19. The width of the interval which they encompass is 0.11-0.23, despite the fact that this, the ratio of the sum of the two lines to Hα, is almost the same as for $R_I^1$.

We were unable to find a direct dependence of the average values of $R_I^1$ and $R_I^2$ for the HII regions on the star formation rate SFR. However, there appears to be a dependence of the ratio of these quantities, $R_I^1/R_I^2$, on SFR (Fig. 8) or, equivalently, of the ratio of the averages over the HII regions of the intensities of the [NII] λ6583 line and the sum of the [SII] λλ6716, 6731 lines. Further studies will show whether this dependence extends to HII regions in other galaxies and can be used as a diagnostic for preliminary estimation of the star formation rates.

The dependence that was obtained is interpreted as a decreasing in $I(\lambda 6583)/I(\lambda 6716+\lambda 6731)$ with increasing rates of star formation SFR of the individual HII regions. It can be said with high probability that for star formation rates $SFR < 0.5 M_\odot$ the emission in the [NII] λ6583 forbidden line of singly ionized nitrogen predominates

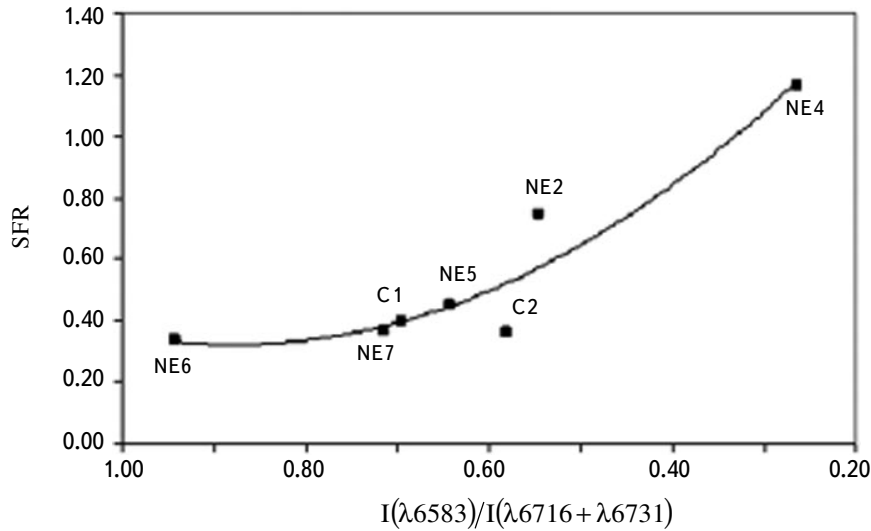

Fig. 8. A plot of the ratio of the intensities of the [NII] λ6583 line and the sum of the [SII] λλ6717, 6731 lines averaged over the HII regions on the abscissa versus the star formation rate SFR (Hα) in units of ($M_\odot$/year) on the ordinate.

TABLE 3. Some Characteristics of the HII Regions According to the MPFS 2007 Data

|  | NE4 | NE2 | C1 | C2 | NE5 | NE6 | NE7 |
|---|---|---|---|---|---|---|---|
| $I_{avg}$ (H$\alpha$)$10^{-16}$ (erg/cm$^2$ s) | 85.3 | 54.0 | 29.8 | 26.4 | 32.8 | 25.0 | 26.9 |
| $R^1_{1\,avg}$ | 0.05 | 0.06 | 0.16 | 0.11 | 0.09 | 0.16 | 0.10 |
| $R^2_{1\,avg}$ | 0.09 | 0.09 | 0.14 | 0.09 | 0.10 | 0.09 | 0.07 |
| $I_{avg}(6583)/I_{avg}(6716+6731)$ | 0.26 | 0.55 | 0.70 | 0.58 | 0.64 | 0.94 | 0.71 |

over that in the forbidden lines of the sulfur [SII] $\lambda\lambda$6716, 6731 doublet.

We note, also, that the emission in the [SII] $\lambda\lambda$6716, 6731 forbidden line is detected only from the HII regions NE4 and NE2 that lie in the upper part of the curve in Fig. 8.

As for the intensity ratio of the lines in the sulfur doublet, $n_e = I(\lambda 6716)/I(\lambda 6731)$, which in the case of photoionization is an indicator of the electron density, if it exceeds 80-100 cm$^{-3}$, then for most of the HII regions it has an average value of 1.2. NE7 is characterized by a slightly lower value, 0.9. NE4 and NE2 are different, with slightly larger values of $n_e$. The average for NE2 is roughly 1.5, and within NE4, it increases from 1.2 at the center to 1.6 at the edge.

## 7. Conclusion

We have presented 3D spectroscopy data for SBS 1202+583 from the 6-m telescope at the SAO of the Russian Academy of Sciences and the 2.6-m telescope at the BAO of the Academy of Sciences of the Armenian Republic that has been acquired as part of a comprehensive study of selected galaxies from the SBS sample. The panoramic MPFS and VAGR spectrographs have been used to obtain data in four overlapping fields, which made it possible, in particular, to analyze the distribution of the emission from SBS 1202+583 in the most intense recombination and forbidden lines in the 490-520 and 600-780 nm ranges. The distribution of the H$\alpha$ emission reveals the existence of nine HII regions, each of which has been studied individually. Six of them are associated with SDSS photometric objects. The distribution of the continuum radiation indicates the existence of a single intensity peak, so that the object under study can be characterized as a unified complex of HII regions.

The presence of seven of the HII objects, including the brightest and largest, in one of the fields, MPFS 2007, ensures uniformity of the data, so that it is possible to estimate the ranges of variation of several characteristics of the HII regions in this study. In particular, the equivalent radii of the HII regions ($R_{eq}$) lie within a range of 84-460 pc, and the rates of star formation (SFR) calculated from the H$\alpha$ luminosity lie within $0.3 - 1.2\, M_\odot/$year. An analysis of the ionized gas forbidden lines reveals a dependence of the ratio of the intensities in the [NII] $\lambda$6583 ionized nitrogen line to the intensity of the sum of the [SII] $\lambda\lambda$6716, 6731 sulfur lines on the star formation rate SFR. It will be more appropriate to reconsider the interpretation of this dependence, which may be related to stratification of the gases, the chemical composition of supernovae, and a variety of other causes, after new data have been collected.

This work was partially based on observations obtained with the 6-m telescope of the Special Astrophysical Observatory of the Russian Academy of Sciences (SAO RAS). The observations were carried out with the financial support of the Ministry of Education and Science of Russian Federation (contracts no. 16.518.11.7073 and 16.552.11.7028).

The Sloan Digital Sky Survey is supported by the Alfred Sloan Foundation with the participation of thirteen institutes directed by the Astrophysical Research Consortium, the U.S. NASA, NSF, and Dept. of Energy, the Japanese Monbukagakusho, and the German Max Planck Society. Its official website is http://www.sdss.org/.